\documentclass[fleqn,usenatbib]{mnras}

\usepackage{newtxtext,newtxmath}

\usepackage[T1]{fontenc}

\DeclareRobustCommand{\VAN}[3]{#2}
\let\VANthebibliography\thebibliography
\def\thebibliography{\DeclareRobustCommand{\VAN}[3]{##3}\VANthebibliography}

\usepackage{graphicx}	
\usepackage{amsmath}	
\usepackage[normalem]{ulem}

\usepackage[dvipsnames]{xcolor}

\title[\sc{dns formation}]{Double neutron star formation via consecutive type II supernova explosions}

\author[V. Fröhlich et al.]{
Viktória Fröhlich,$^{1,2,3}$\thanks{E-mail: frohlich.viktoria@csfk.org}
Zsolt Regály,$^{1,2}$
József Vinkó$^{1,2,4,5}$
\\
$^{1}$Konkoly Observatory, Research Centre for Astronomy and Earth Science, Konkoly-Thege Mikl\'os 15-17, 1121, Budapest, Hungary\\
$^{2}$CSFK, MTA Centre of Excellence, Budapest, Konkoly Thege Miklós út 15-17., H-1121, Hungary\\
$^{3}$E\"otv\"os Lor\'and University, P\'azm\'any P\'eter s\'et\'any 1/A, 1117 Budapest, Hungary\\
$^{4}$ELTE E\"otv\"os Lor\'and University, Institute of Physics, P\'azm\'any P\'eter s\'et\'any 1/A, Budapest, 1117 Hungary\\
$^{5}$Institute of Physics, University of Szeged, D\'om t\'er 9, Szeged, 6720, Hungary
}

\date{Accepted XXX. Received YYY; in original form ZZZ}

\pubyear{2023}

\begin{document}
\label{firstpage}
\pagerange{\pageref{firstpage}--\pageref{lastpage}}
\maketitle

\begin{abstract}

Since the discovery of the first double neutron star (DNS) system, the number of these exotic binaries has reached
fifteen. Here we investigate a channel of DNS formation in binary systems with components above the mass limit of type II supernova explosion (SN II), i.e. $8M_\odot$. We apply a spherically symmetric homologous envelope expansion model to account for mass loss, and follow the dynamical evolution of the system numerically with a high-precision integrator.  The first SN occurs in a binary system whose orbital parameters are pre-defined, then, the homologous expansion model is applied again in the newly formed system. Analysing 1 658 880 models we find that DNS formation via subsequent SN II explosions requires a fine-tuning of the initial parameters. Our model can explain DNS systems with a separation greater than 2.95 au. The eccentricity of the DNS systems spans a wide range thanks to the orbital circularisation effect due to the second SN II explosion. The eccentricity of the DNS is sensitive to the initial eccentricity of the binary progenitor and the orbital position of the system preceding the second explosion. In agreement with the majority of the observations of DNS systems, we find the system centre–of mass velocities to be less than $60~\mathrm{km~s^{-1}}$. Neutron stars that become unbound in either explosion gain a peculiar velocity in the range of $0.02-240~\mathrm{km~s^{-1}}$. In our model, the formation of tight DNS systems requires a post-explosion orbit-shrinking mechanism, possibly driven by the ejected envelopes.

\end{abstract}

\begin{keywords}
binaries: general -- stars: evolution -- stars: mass-loss -- stars: neutron -- supernovae: general
\end{keywords}



\section{Introduction}
\label{sec:intro}

The number of high-mass binaries in the Galaxy is rather moderate. 
In the catalogues of \citet{torres-etal-2010, eker-etal-2014}, and \citet{southworth-2015} only 5-11 per cent of the observed binaries have two components above the type II supernova (SN~II) mass limit ($8M_\odot$).
Furthermore, there are no binaries with both components heavier than $8M_\odot$ in the survey of \citet{soderhjelm-1999} and in the Gaia catalogue of binaries with measured masses \citep{gaia-2022}.
In the case of the Gaia mission, however, we must bear in mind that the time span of DR3 (33 months) is barely sufficient to obtain high quality data on binary systems or to infer statistical properties, and that both astrometric and photometric biases are prominent in the data set \citep{gaia-biases}. 
It should also be noted that many high-mass, short-lived OB stars spend their entire lives in the Galactic disk, where they are heavily obscured by dust and are difficult to detect in large numbers by the Gaia instrument.

The above-mentioned massive binaries can, in principle, provide a channel for the formation of double neutron star (DNS) systems.
If both stars are massive enough to end their lives in core-collapse supernovae (SN), the binary may evolve into a DNS system. 
The most widely accepted theory of DNS formation is that mass transfer takes place during the lifetime of a binary (see e.g. \citealp{bhattacharya-vdheuvel-1991, tauris-vdheuvel-2006, tauris-etal-2017} for a summary).
Stars whose envelopes are stripped away by mass transfer or stellar winds explode as SN Ib or Ic (e.g. \citealp{eldridge-etal-2008}). 
When such low-mass helium stars explode, the ejecta mass can be as low as $0.1~M_\odot$ \citep{tauris-etal-2015}. 
Consecutive SN Ib/Ic explosions are considered to be the main formation channel for DNS systems that merge within a Hubble time. 
According to this canonical scenario of binary interactions, the primary neutron star is eventually engulfed by the envelope of the other star.
However, \citet{brown-1995} and \citet{dewi-etal-2006} suggested that the neutron star can avoid engulfment if the components leave the main sequence before evolving into contact and the pulsars are born from the explosions of helium stars.
These models can only explain DNS systems whose progenitors are of equal mass, as was emphasised by \citet{tauris-etal-2017}.
There is also a chance that the components of a binary system do not fill their Roche lobes. 
It is therefore possible that the stars do not exchange significant mass before exploding as Type II supernovae (SN~II).

Observational evidence for DNS systems has been well established.
If the components of a DNS system orbit close enough to merge, the intensity and frequency of the emitted gravitational waves (GW) will eventually fall within the sensitive range of GW observatories, and the merging event can be detected (see, for example, \citealp{abott-etal-2017}).
However, current detectors (LIGO, VIRGO, GEO600 and KAGRA) are not sensitive enough to observe DNS systems until seconds before the merger.

Gamma-ray bursts (GRBs) provide further evidence for the existence of DNS systems.
Short GRBs are formed by the merger of compact objects: two NSs or one NS and a stellar-mass back hole.
The short GRB 170817A has been directly linked to the first NS merger detected in the form of a GW \citep{abott-etal-2017, goldstein-etal-2017}.
NS-NS mergers have also been theorised to produce kilonovae \citep{li-paczynski-1998}.
Indeed, the kilonova 130603B has been associated with a short GRB and thus with a merging DNS system \citep{tanvir-etal-2013}. 
Furthermore, the first GW-emitting NS merger has also been associated with a kilonova event \citep{arcavi-etal-2017, coulter-etal-2017, lipunov-etal-2017, tanvir-etal-2017, soares-santos-2017, valenti-etal-2017}.
It has also been suggested by \citet{norris-2002, norris-bonnel-2006} and \citet{gehrels-etal-2006} that a subset of short GRBs - extended emission GRBs - also form in mergers and thus may be a tracer of DNS systems.
Long GRBs are usually associated with the core collapse of massive, rapidly rotating, low-metallicity stars that form black holes \citep{galama-etal-1998, woosley-bloom-2006}.
However, long GRB events have also been associated with kilonovae \citep{rastinejad-etal-2022} and with the merger of compact objects \citep{troja-etal-2022}.

In addition to the above-mentioned observational channels, there is a slight chance that the jets emitted from the magnetic poles of the neutron stars periodically point towards Earth.
In such a lucky scenario, DNS systems can be observed as double radio pulsars \citep{lorimer-kramer-2004}.

To date, fifteen DNS systems have been discovered, the detailed descriptions of which can be found in the review by \citet{tauris-etal-2017}.
These systems show a wide range of orbital eccentricities (0.085-0.83).
However, they are all close binaries with orbital periods of $\leq$~45 days.
The system velocities of the binaries are also diverse, ranging from 28 to 240~$\mathrm{kms^{-1}}$.
Discoveries of these systems are either based on pulsar timing \citep{fonseca-etal-2014} or observing an eclipse \citep{breton-etal-2008}.
Some globular clusters also host DNS systems \citep{phinney-sigurdsson-1991, prince-etal-1991, grindlay-etal-2006, lynch-etal-2012, verbunt-freire-2014}. 

In  globular clusters or galactic bulges where the spatial density of the remnants of massive stars is high, capture of a NS by another NS may be most effective formation channel of DNS systems.
However, due to the relative scarcity of stellar remnants, capture in any other environment is highly ineffective.
In this case, we have to assume that two consecutive SN explosions occur within the same binary system.

Models investigating the formation of DNS systems have taken both an analytical \citep{flannery-vdheuvel-1975, hills-1983, kalogera-1996, tauris-takens-1997} and a numerical approach \citep{kornilov-lipunov-1983, dewey-cordes-1987, brandt-podsiadlowski-1995}. 
However, only a few of these models \citep{dewey-cordes-1987, brandt-podsiadlowski-1995} study the explosion of both stars.
The mass loss of the star is not modelled in any of the above works.
Instead, the effects of NS kicks on the orbital elements and the peculiar velocity of the systems are studied, assuming an instantaneous, asymmetric SN explosion.
These models are able to explain the high peculiar velocity of solitary PSRs and the tightness of the observed DNS systems by fine-tuning the direction and magnitude of the NS kicks.
    
In this study, we investigate a model in which subsequent SN~II explosions are assumed to be a plausible channel for the formation of DNS systems.
Unlike any previous DNS formation study, mass loss is modelled with a simple but physically more accurate envelope expansion.
As a result of the heavy mass loss during an SN~II explosion, the orbital stability of a system might break down as was first demonstrated by \citet{hadjidemetriou-1966} and recently by the pioneering work of \citet{veras-etal-2011}.
After explosion, the orbital eccentricity can often grow above unity, thus the stars become unbound and gain high peculiar velocities.
However, in some cases the binary remains bound forming a DNS system.
The variable-mass two-body problem in a binary star system, utilising a homologous SN~II expansion model, was first explored numerically by \citet{regaly-etal-2022}.
Our work showed that double star systems have an approximately 14 per cent chance of surviving a single SN~II explosion, providing a possible channel for DNS formation.
In this study, we extend our previous work to binary star systems in which both stars explode, i.e. both their masses are above the SN~II limit ($\geq8M_\odot$).
The two explosions are modelled with the spherically symmetric homologous expansion model presented in \citet{regaly-etal-2022}.
The orbital elements and velocities of the components are monitored through both explosions and beyond.
We discuss in detail a DNS formation scenario that can explain the formation of wide–orbit DNS systems.

This paper is structured as follows.
After the Introduction, the summary of the applied numerical model is given in Section~\ref{sec:methods}, along with the parameter regimes explored in our study.
Our results and analysis are presented in Section~\ref{sec:results}.
In Section~\ref{sec:discussion}, we discuss the properties and the accompanying phenomena of system evolution.
Finally, our results and conclusions are summarised in Section~\ref{sec:conclusions}.

\begin{figure*}
    \centering
    \includegraphics[width=0.9\textwidth]{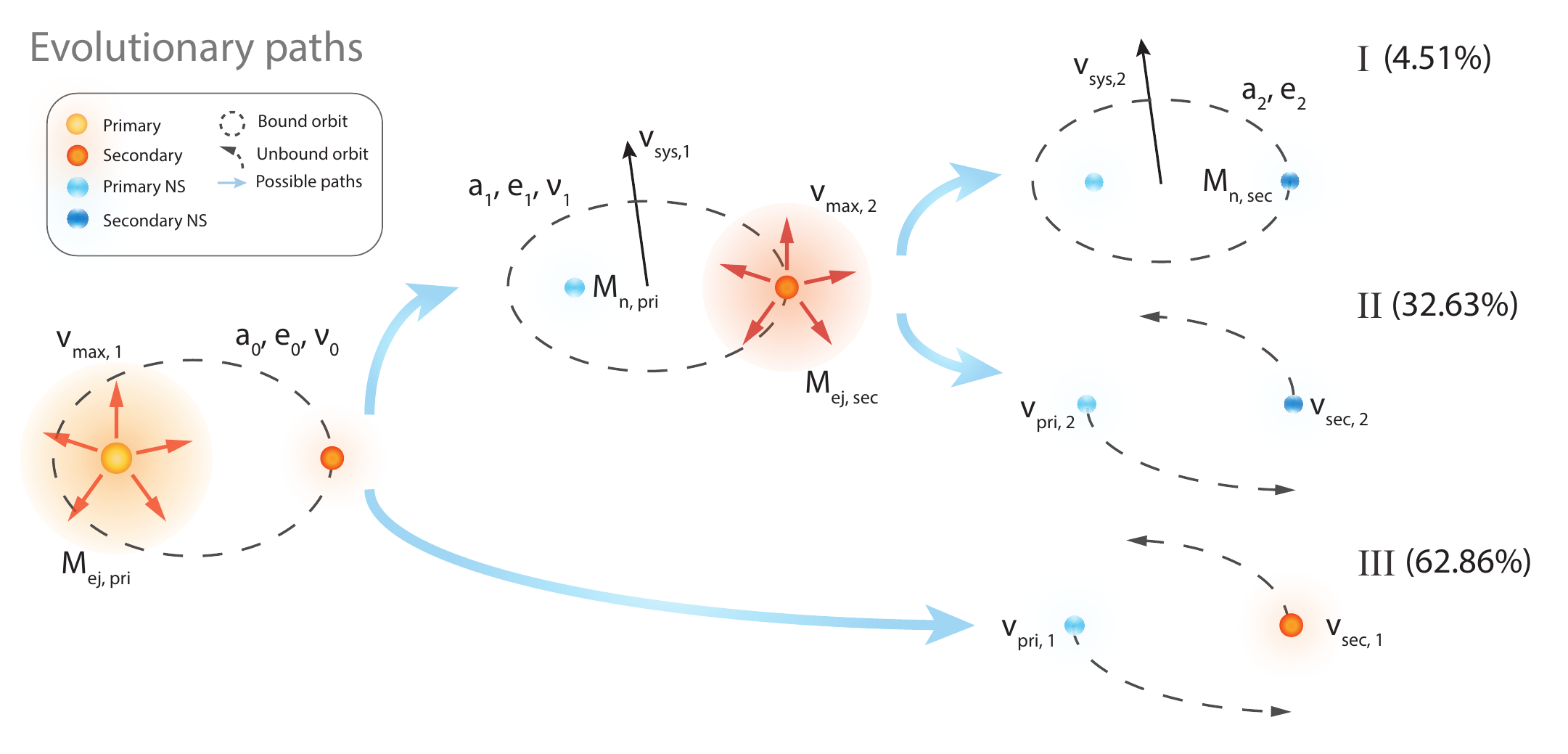}
    \caption{Visual representation of the investigated parameters and monitored values, along with the possible evolutionary paths of the modelled systems.  
    A system that follows path I stays bound after both SN~II explosions and evolves into a DNS system. 
    A binary that dissociates after the second or the first explosion follows path II or III, respectively.
    Percentages show the occurrence rates of the paths in our model set.
    The primary SN~II progenitor star is yellow, while the smaller secondary SN~II progenitor is orange.
    The primary NS is light blue, and the secondary NS is dark blue.}
    \label{fig:models}
\end{figure*}

\section{Numerical methods}
\label{sec:methods}

\subsection{Variable mass two–body approach}

In this study, we model the orbital dynamics of a binary star system where both components undergo subsequent SN~II explosions.
The mass of the first exploding star, which is considered to be the primary component of the system (denoted by the subscript pri throughout this study) is initially larger than that of the companion.
Note that the star that undergoes the second explosion -- considered to be the secondary and denoted by the subscript sec -- has a larger mass compared to the primary at the onset of the second explosion.
To model the orbital evolution of these systems, we numerically solve the equations of motion for the varying mass two-body problem, which are
\begin{equation}
    \ddot{\textbf{r}}_\mathrm{pri}=-M_\mathrm{in,pri} \ G \ \frac{\textbf{r}_\mathrm{pri}}{r_\mathrm{pri}^{3}},
    \label{eq:mo1}
\end{equation}
\begin{equation}
    \ddot{\textbf{r}}_\mathrm{sec}=-M_\mathrm{in,sec} \ G \ \frac{\textbf{r}_\mathrm{sec}}{r_\mathrm{sec}^{3}},
    \label{eq:mo2}
\end{equation}
where $G$ is the gravitational constant, while $\textbf{r}_\mathrm{pri}$ and $\textbf{r}_\mathrm{sec}$ are the position vector of the primary and secondary components in the inertial frame, respectively. 
The mass inside the primary's orbit is $M_{\mathrm{in,pri}}$, and the mass inside the secondary's orbit is $M_{\mathrm{in,sec}}$.
Before the SN explosions, $M_{\mathrm{in,pri}}=M_{\mathrm{sec}}$ and $M_{\mathrm{in,sec}}=M_{\mathrm{ej}}+M_{\mathrm{n}}$
During the SN\,II explosions $M_{\mathrm{in,pri}}$ and $M_{\mathrm{in,sec}}$ are subject to change in time, however, $M_{\mathrm{in,pri}}$ is constant during the first, and $M_{\mathrm{in,sec}}$ is constant during the second explosion.

To derive the mass residing inside the orbits of each component, we assume a spherically symmetric homologous expansion of the SN envelopes (presented e.g. in \citealp{arnett-1980, vinko04, bw17} and further developed in \citealp{regaly-etal-2022}).
The homologous nature of the expansion means that: 
1) the velocity of each ejected layer is a linear function of distance from the centre of the SN; 
2) the velocity of the outermost layer is time-invariant; and 
3) the density profile of the ejected material is also independent of the elapsed time. 
We assume that the density profiles of the two stars are similar: they both contain a constant-density core extending up to a fractional radius of $x_{\mathrm{c}} = r_{\mathrm{c}} / R_{0}$, where $r_{\mathrm{c}}$ is the initial radius of the core and $R_{0}$ is the initial radius of the progenitors.
$R_0$ is considered to be $500R_\odot$ throughout this study.
The outer envelope has a power-law density profile described by the exponent $n$, which is fixed as $n=7$ throughout this study based on \citet{hatano-etal-1999}.
We have investigated the case of $n = 2$ and have not found any qualitative differences, but a more detailed investigation will be left for a future study.

When applying this homologous expansion model, three cases should be distinguished.
In the first case, all mass still resides within the orbit of the non-exploding star, i.e. $M_{\mathrm{in}}=M_{\mathrm{ej}}+M_{\mathrm{n}}$, where $M_{\mathrm{ej}}$ is the mass of the envelope lost in the SN explosion, and $M_{\mathrm{n}}$ is the mass of the remnant neutron star.
In the second case, the co-moving distance coordinate of the companion is larger than that of the core,  $x_\mathrm{comp}>x_\mathrm{c}$, and
\begin{align}
    \begin{split}
        M_\mathrm{in}&=M_\mathrm{n} + M_\mathrm{c}+ 4 \pi \rho_\mathrm{0} R_\mathrm{0}^3 x_\mathrm{c}^n \int_{x_\mathrm{c}}^{x_\mathrm{comp}} x^{2-n} dx =\\
        &= M_\mathrm{n} + M_\mathrm{c} \left [ 1+ \frac{3}{n-3} \left ( 1- \left ( \frac{x_\mathrm{comp}}{x_\mathrm{c}} \right )^{3-n} \right ) \right ].    
    \end{split}
    \label{eq:min_xp>xc}
\end{align}
In the third case, $x_\mathrm{comp}<x_\mathrm{c}$, thus
\begin{align}
    \begin{split}
    M_\mathrm{in}&=M_\mathrm{n}+ 4 \pi \rho_\mathrm{0} R_\mathrm{0}^3  \int_{0}^{x_\mathrm{comp}} x^2 dx =\\
    &= M_\mathrm{n} + M_\mathrm{c} \left ( \frac{x_\mathrm{comp}}{x_\mathrm{c}} \right )^3.
    \end{split}
    \label{eq:min_xp<xc}
\end{align}
In the above equations, $M_{\mathrm{c}}$ is the mass of the core of the SN ejecta having a constant density $\rho_0$.

The equations of motion given in Eqs.\,(\ref{eq:mo1}) and (\ref{eq:mo2}) are solved numerically in two dimensions using the Runge-Kutta method discussed in \citet{regaly-etal-2022}.
The integration time for each explosion is 5 million days (about 13,700\,yr).
This allows the monitored physical quantities, namely the semi-major axis ($a$), the eccentricity ($e$), and the orbital velocities ($v$) to level off at a constant value by the end of the simulations.
If a given model remains bound after the first explosion, i.e. the eccentricity of the orbits is below unity, the homologous expansion is applied again to model the explosion of the secondary star.
Orbital parameters and velocities of components are calculated at the last time step of the integration for both explosions. 
If a system dissociates after the first explosion, the integration is stopped and the explosion of the secondary is not modelled.
Even though the secondary in this case gains high peculiar velocity, its spherically symmetric explosion will not alter its spatial motion, so no further integration is necessary.

\subsection{Investigated parameter regimes}

In order to map out possible DNS formation scenarios, we study different initial conditions (stellar masses and orbital parameters).
The masses of the binary components are $8-25M_\odot$, which is in between the minimum mass required for a SN~II explosion and the mass limit of black hole formation \citep{fryer-1999, mao-etal-2021}.
Since envelope stripping and stellar winds cannot be described with the homologous expansion model, we neglect their effects throughout this study.
This means that $M_\mathrm{ej,pri}+M_\mathrm{n,pri}$ and $M_\mathrm{ej,sec}+M_\mathrm{n,sec}$ are the initial masses of the binary components.
The minimum NS mass is constrained by the Chandrasekhar limit, and the upper mass limit is set by both theory and observations \citep{kalogera-baym, clark-etal}.
Note that according to observations, the mass of NSs is likely to be 1.1-2.35 $M_\odot$ \citep{tauris-etal-2015, romani-etal-2022}, and is notably low in DNS binaries (1.17-1.56 $M_\odot$). 
NSs with masses close to the Tollman-Oppenheimer-Volkoff limit are a rare sight.
The velocities of the outermost expanding shell for both components ($v_{\mathrm{max,1}}$ and $v_{\mathrm{max,2}}$) are in the range of 1000-10000~$\mathrm{km~s^{-1}}$ based on \citet{hamuy-pinto}.

Initial orbital elements of the system are denoted by the subscript 0.
Subscripts 1 and 2 denote the orbital elements, the velocity of the system centre-of-mass (hereinafter referred to as system velocity) and the component velocities after the first and second explosions, respectively.
If a system remains bound, its system velocity, $v_{\mathrm{sys}}$, is also calculated.
If the system dissociates during either explosion, the peculiar velocities of both components, $v_{\mathrm{pri}}$ and $v_{\mathrm{sec}}$, are monitored.
The true anomaly, ($\nu$, measured at the moment of explosion) of the system can be chosen arbitrarily before each explosion.

The explored parameter regimes are given in Table~\ref{tab:params}, giving a total of 1~658~880 models.
For a visual summary of the used and monitored parameters and possible evolutionary paths of the binary system, see Figure~\ref{fig:models}.
The occurrence rate of each evolutionary path is also noted in Figure~\ref{fig:models}.

\begin{table}
\begin{center}
\begin{tabular}{ll}
\hline
Parameter & Values \\
\hline
$a_\mathrm{0}$  & $R_0$, 10~au, 25~au, 50~au, 100~au \\ 
$e_\mathrm{0}$  & $0.0, 0.2, 0.4, 0.8$  \\ 
$\nu_{\mathrm{0}}$  & $0.0,\pi/4,\pi/2,3\pi/4,\pi,5\pi/4,6\pi/4,7\pi/4$  \\ 
$\nu_{\mathrm{1}}$  & $0.0,\pi/4,\pi/2,3\pi/4,\pi,5\pi/4,6\pi/4,7\pi/4$  \\ 
$v_\mathrm{max,1}$  & 1000~$\mathrm{km~s^{-1}}$, 5000~$\mathrm{km~s^{-1}}$, 10000~$\mathrm{km~s^{-1}}$ \\ 
$v_\mathrm{max,2}$  & 1000~$\mathrm{km~s^{-1}}$, 5000~$\mathrm{km~s^{-1}}$, 10000~$\mathrm{km~s^{-1}}$ \\ 
$M_\mathrm{ej,pri}$  & $6.5M_\odot, 12M_\odot, 15M_\odot, 22M_\odot^\dag$  \\ 
$M_\mathrm{ej,sec}$  & $6.5M_\odot, 12M_\odot, 15M_\odot, 22M_\odot^\dag$  \\ 
$M_\mathrm{n,pri}$  & $1.5M_\odot, 2M_\odot, 3M_\odot^\dag$  \\ 
$M_\mathrm{n,sec}$  & $1.5M_\odot, 2M_\odot, 3M_\odot^\dag$  \\  \hline
\end{tabular}
\end{center}
\label{tab:params}
\caption{Investigated system parameters. $^\dag$Note that only those models are analysed, where $M_{\mathrm{ej,pri}}+M_{\mathrm{n,pri}} \geq M_{\mathrm{ej,sec}}+M_{\mathrm{n,sec}}$.}
\end{table}

\section{Results}
\label{sec:results}

\subsection{Bound systems}

In this section, we analyse models where the eccentricity of the binary system remains below unity after both the first and the second SN~II explosion, i.e., follows the evolutionary path I (see Figure~\ref{fig:models}).
In order to form a DNS system, it is necessary that the binary remains bound after both explosions.
In general, we find that the higher the ejecta mass, the lower the number of systems that stay bound.
This trend is clearly identifiable in the case of the second explosion.
Note that an opposite trend is found in the first explosion, which can be explained by a selection bias of the examined parameter regimes:
models where $M_{\mathrm{sec}}>Mp_{\mathrm{pri}}$ are eliminated from our analysis, which is responsible for the limited number of low ejecta mass models.
The mass of the neutron stars does not affect the orbital elements and velocities after either explosion.
However, the number of bound models increases with the mass of the primary NS, $M_{\mathrm{n,pri}}$, while $M_{\mathrm{n,sec}}$ has no effect in this regard.
The number of bound systems is inversely proportional to the maximum velocity of the SN ejecta. 
Interestingly, $v_{\mathrm{max,1}}$ has a stronger effect on the number of bound systems than $v_{\mathrm{max,2}}$.
The likelihood of a system surviving both explosions is significantly higher (by 20-25 per cent) for compact systems, which is in agreement with the findings of \citet{tauris-etal-2017}.
The orbital eccentricity of the pre-SN system also affects the orbital stability.
Interestingly, in the first explosion, orbital eccentricity weakens stability, while in the second explosion, this trend is reversed.

\begin{figure*}
    \centering
    \includegraphics[width=0.9\textwidth]{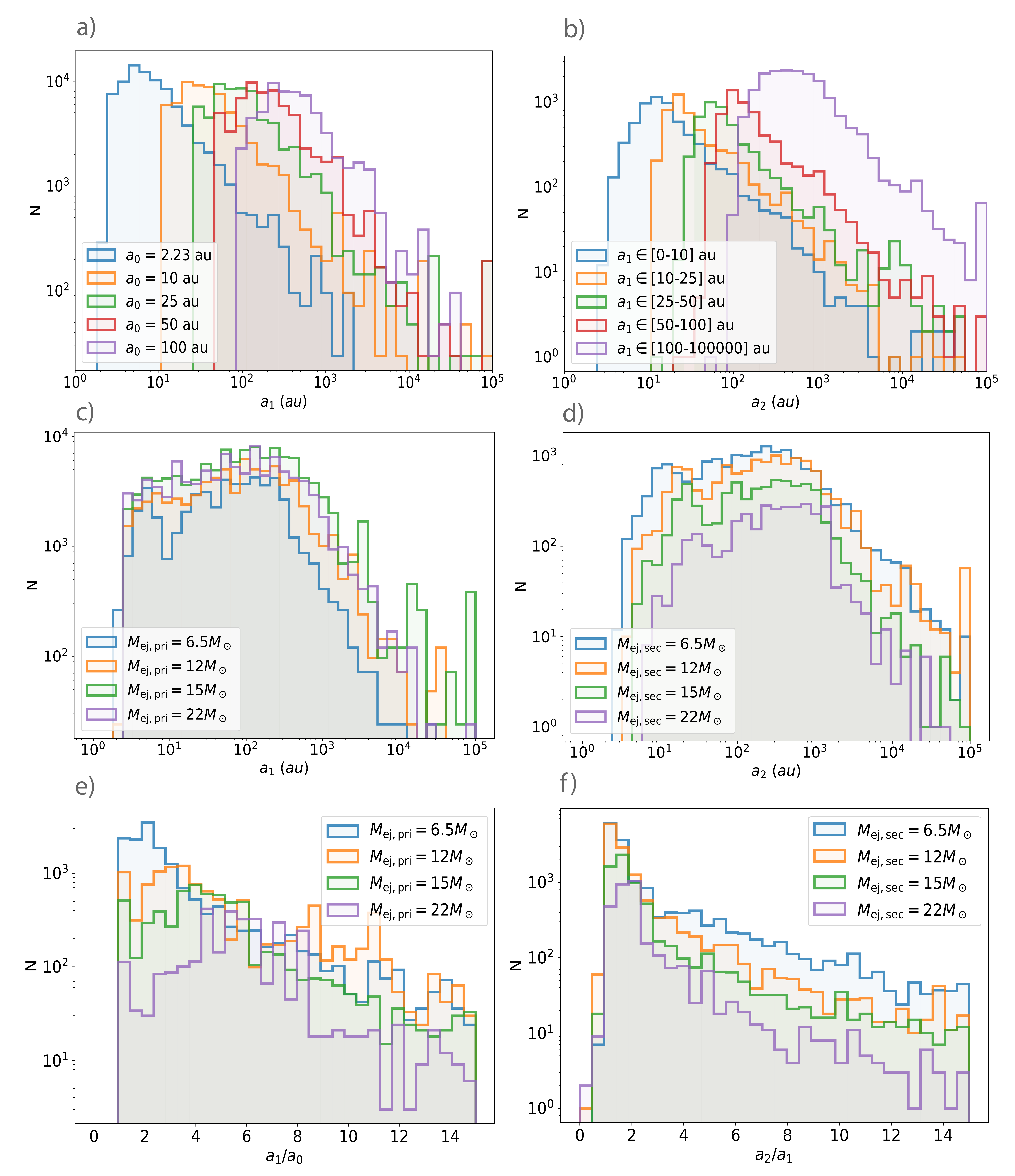}
    \caption{Distributions of the semi-major axes of bound systems after the first (panel a) and second (panel b) explosions.
    Colours represent initial semi-major axis values on panel a, and arbitrarily selected semi-major axis ranges on panel b.
    Panels c and d are the same as panels a and b, but with colours indicating different $M_{\mathrm{ej,pri}}$ values on panel c, and different $M_{\mathrm{ej,sec}}$ values on panel d.
    Logarithmic scales are used for clarity on panels a–d.
    Panels e and f show the distributions of the orbital widening caused by each SN explosion, $a_1/a_0$ and $a_2/a_1$, with the colour coding matching those of panels c and d.}
    \label{fig:a}
\end{figure*}

Figure~\ref{fig:a} shows the histograms of the semi-major axes of the NS-SN progenitors formed in the first explosion, and that of the DNS systems formed in the second explosion.
One can see that both explosions widen the orbit of the binary.
It is evident that by increasing the pre-explosion separation of the system ($a_0$ or $a_1$), the width of the distribution of the final semi-major axis increases.
It is also visible on panel b of Figure~\ref{fig:a}, that the formation rate of very wide orbit binaries (larger than 100~au, represented with purple colour) is larger in the second explosion.
DNS systems have a minimum separation of $a_2=2.95$~au.

On panels c and d of Figure~\ref{fig:a} one can see that the distributions of the semi-major axes overlap, meaning that the separation of the NS-SN progenitor systems, $a_1$, and that of the DNS systems, $a_2$, is independent of the ejecta mass.
The semi-major axis of the systems generally increases, as shown on panels e and f of Figure~\ref{fig:a}.
Interestingly, we find that the separation of the binaries does not grow in the second explosion if $a_1>1000~\mathrm{au}$.
For 204 models, the orbit of the binary shrinks in the second explosion ($a_2<a_1$), and the ratio of the semi-major axes, $a_2/a_1$, has a minimum of 0.3.
The minimum separation of these shrinking binaries is $a_2=11.5~\mathrm{au}$. 
Shrinking of the binary orbit requires that the expansion velocity of the second SN is low ($v_{\mathrm{max,2}}=1000$~$\mathrm{km~s^{-1}}$), and the eccentric ($e_1>0.88$) binary is exactly at apocenter ($\nu_1=\pi$) at the moment of explosion.
Although the values of the semi-major axes are independent of the ejecta mass, the most probable value of the relative change in the semi-major axis in the second explosion shifts towards larger values with increasing $M_{\mathrm{ej,sec}}$.
We find that $1<a_1/a_0<15$, while $0.3<a_2/a_1<15$.
For 51 per cent of models, $a_1/a_0>3$, however, $a_2/a_1>3$ is only valid for 19 per cent of the models.

\begin{figure*}
    \centering
    \includegraphics[width=0.9\textwidth]{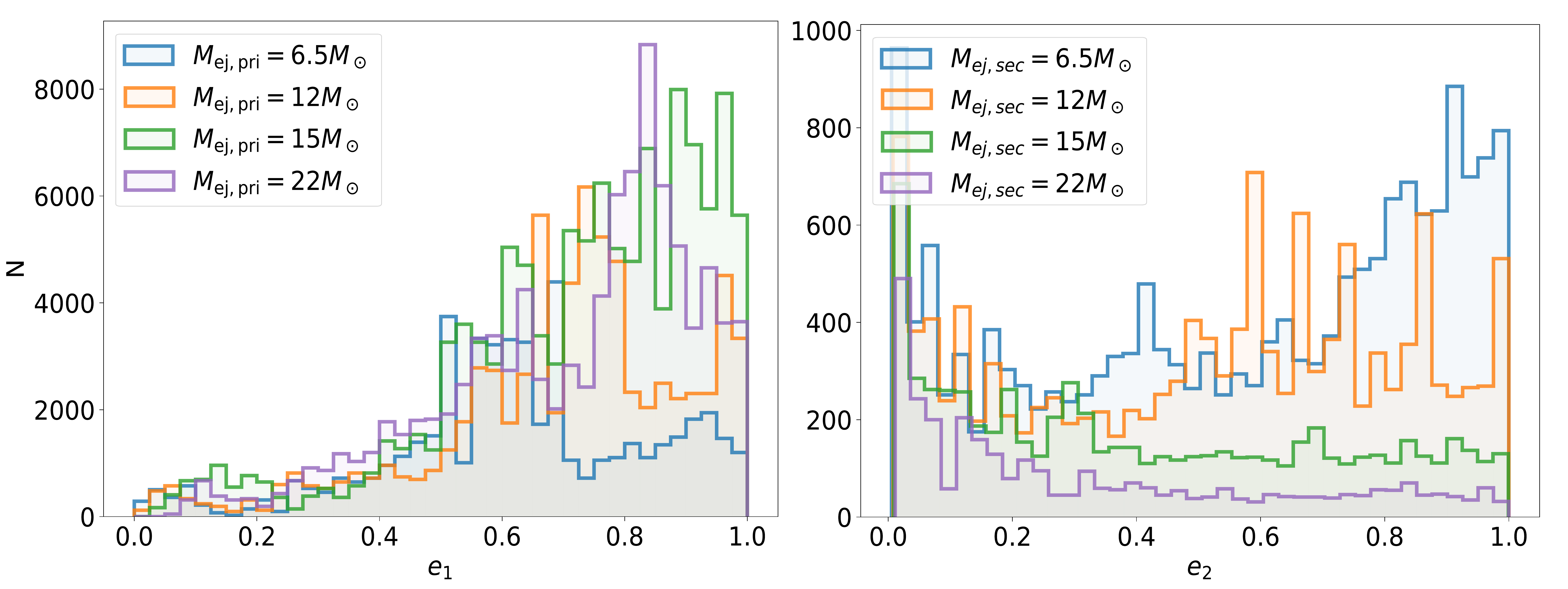}
    \caption{Eccentricities of bound systems after the first (left) and second (right) explosions.
    Colours represent different $M_{\mathrm{ej,pri}}$ values on the left and different $M_{\mathrm{ej,sec}}$ values on the right.}
    \label{fig:e}
\end{figure*}

In Figure~\ref{fig:e} one can see the distributions of the eccentricities of the bound systems.
The eccentricity of bound systems after the first explosion, $e_1$, is found to be independent of the mass ejected in the first explosion, $M_{\mathrm{ej,pri}}$, as can be seen in the left panel.
On the contrary, the smaller the $M_{\mathrm{ej,sec}}$, the larger the eccentricity after the second explosion, $e_2$, as can be seen in the right panel of Figure~\ref{fig:e}.
The eccentricity of 83 per cent of models grows during the first explosion.
However, the eccentricity decreases for almost all (75 per cent) of the models where $e_0=0.8$.
The vast majority of these models (80 per cent) are at the apocentre of the orbit at the moment of the first explosion.
These findings are in agreement with our previous results on planetary-mass companions in \citet{regaly-etal-2022}.
The number of systems that remain bound with $e_1<0.4$ is negligible (9 per cent) within the simulated parameter regimes.
Compared to $e_1$, $e_2$ shows a much more homogeneous distribution in the examined parameter regimes, which implies that the second explosion can circularise the binary systems (see the right panel of Figure~\ref{fig:e}).
Circularisation occurs for $e_1\gtrsim0.4$, which homogenises the distribution of $e_2$.
If $e_1\lesssim0.4$, the orbital eccentricity increases, and the most probable value shifts toward higher values.

\begin{figure*}
    \centering
    \includegraphics[width=0.9\textwidth]{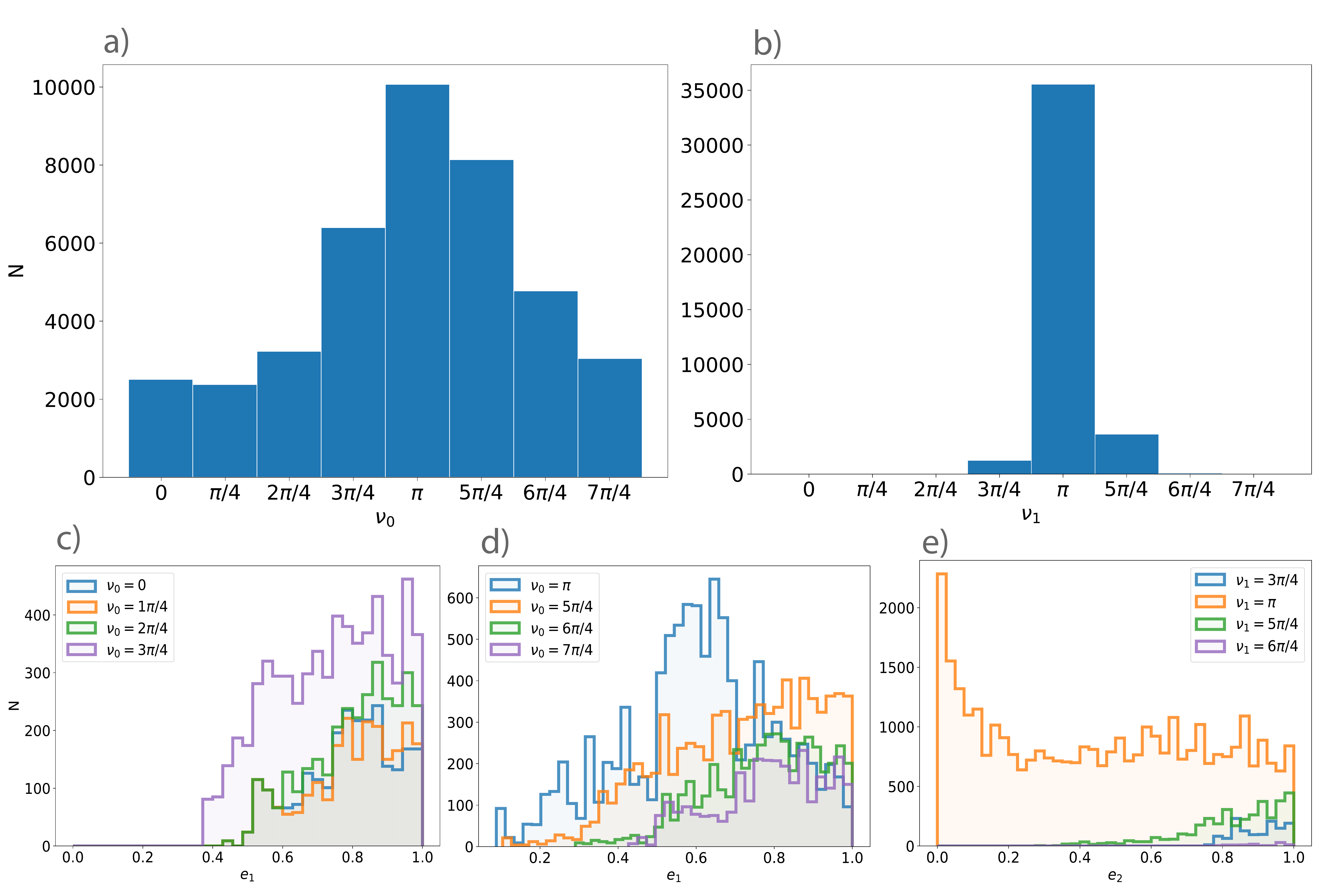}
    \caption{
    Distribution of the number of bound systems as a function of the initial true anomaly in the first (panel a) and second explosions (panel b).
    Eccentricity distributions of bound systems can be seen on panels c-d and e.
    Panels c and d both show the same distribution, for distinct $\nu_0$ ranges, $0-3\pi/4$ and $\pi-7\pi/4$, respectively.}
    \label{fig:nu}
\end{figure*}

Figure~\ref{fig:nu}. shows the distribution of the true anomalies of the companion at the moment of the first (panel a) and second explosion (panel b) for bound systems.
Systems can remain bound after the first explosion with an arbitrary $\nu_0$ true anomaly value, as can be seen on panel a.
The most favourable value for the NS–SN progenitor binary stability is $\nu_0 \sim \pi$, i.e. when the secondary is at the apocentre when the primary star explodes.
In the second explosion only those systems can remain bound after the second explosion whose $\nu_1\approx \pi$ (see panel b).
This can be explained by the fact that the mass of the primary neutron star is small compared to the exploding secondary.
The distribution of the number of bound models is asymmetric for both cases and is pushed towards values greater than $\pi$.
This is due to the orbital movement of the bodies during the first phase of the SN~II explosion, as discussed in detail by \citet{regaly-etal-2022}.
The panels c-e of Figure~\ref{fig:nu} exhibit the distributions of the eccentricities $e_1$ and $e_2$, colour-coded according to the pre-explosion true anomalies $\nu_1$ and $\nu_2$, respectively.
For clarity, the distributions of $e_1$ are presented in two panels (c and d).
We find that the closer the stars are to the apocentre of the orbit, the wider the distributions of the eccentricities (the minimum can reach $e_1<0.1$, see panel d), while if the stars are close to the pericentre, the distributions collapse and their minimum shifts to a larger value (panel c).
This phenomenon can also be observed during the second explosion, as seen in panel e.

\begin{figure*}
    \centering
    \includegraphics[width=0.9\textwidth]{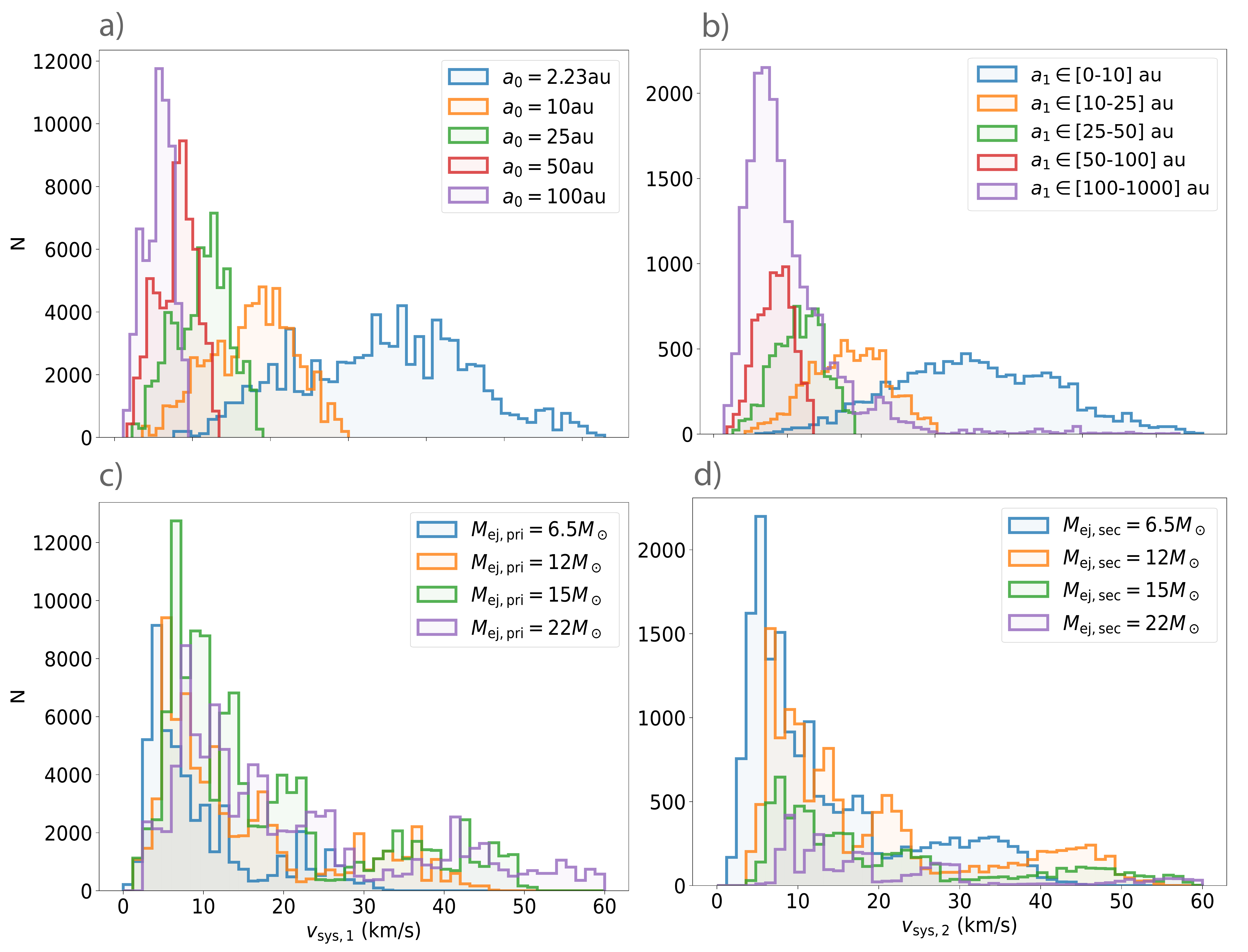}
    \caption{
    Bound systems' velocities gained in the first (panel a) and second (panel b) explosions.
    Colours represent different values and ranges of the pre-SN separation of the binaries.
    Note that the semi-major axes after the first explosion are an outcome of the simulations and are not pre-defined values.
    Panels c and d are similar to panels a and b, however, colours indicate  $M_{\mathrm{ej,pri}}$ and $M_{\mathrm{ej,sec}}$ values.} 
    \label{fig:vsys}
\end{figure*}

For the bound systems the centre of mass gains a non-zero peculiar velocity regardless of the spherically symmetric nature of the homologous expansion models.
This can be explained by the fact that the expanding shell inherits the orbital velocity of the exploding star at the moment of explosion, while the binary gains equal but opposite momentum due to the conservation of momentum.
system velocity is smaller than 60~$\mathrm{km~s^{-1}}$, and most likely falls in the range of 5-10~$\mathrm{km~s^{-1}}$ for both explosions.
The distributions of the system velocities are shown in Figure~\ref{fig:vsys}.
Panels a and b are colour-coded with the pre-explosion separation, while the colors in panels c and d represent the ejecta masses.
It is visible in panels a-b of Figure~\ref{fig:vsys} that the smaller the separation of the system at the moment of explosion, the higher the system velocity can be.
This tendency is clear in the case of the first explosion ($v_{\mathrm{sys,1}}$, panel a).
Note that in panel b the distributions are colour coded with ranges of $a_1$ because the separation before the second explosion is not pre-defined, but is an outcome of the simulations. 
The system velocity of bound models is independent of the ejecta mass in the first explosion (panel c of Figure~\ref{fig:vsys}), however, it is weakly dependent on $M_{\mathrm{ej,sec}}$ (panel d).
In the second explosion, the smaller the ejecta mass, the smaller the velocity of the system.

\subsection{Unbound systems}

\begin{figure*}
    \centering
    \includegraphics[width=0.9\textwidth]{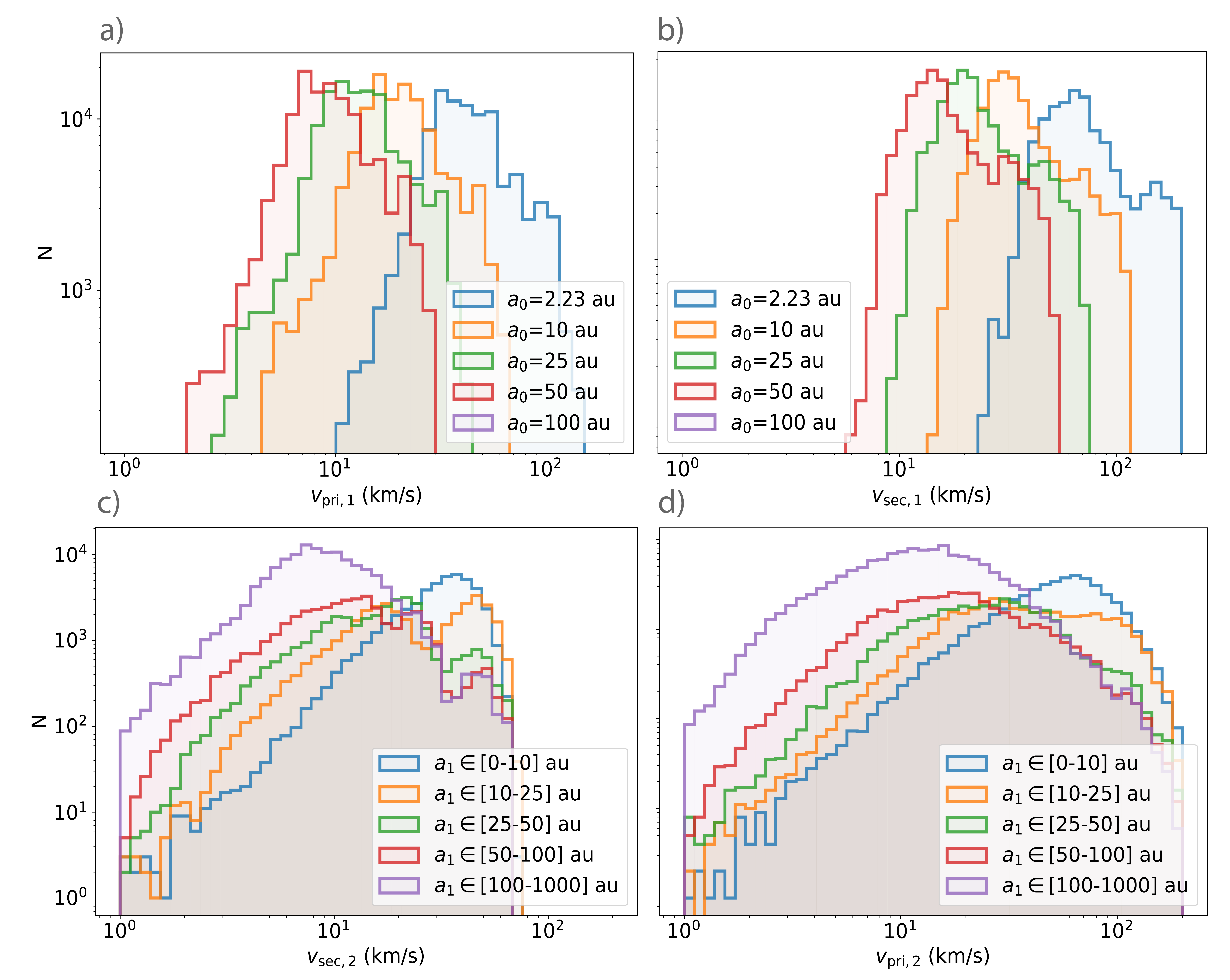}
    \caption{Distributions of the component velocities of unbound systems.
    In panels a and b one can see the velocities of the primary NS ($v_{\mathrm{pri,1}}$) and the SN progenitor ($v_{\mathrm{sec,1}}$) if the system dissociates in the first explosion.
    Panels c and d show peculiar velocities of the primary NS ($v_{\mathrm{pri,2}}$) and secondary NS ($v_{\mathrm{sec,2}}$) gained in the second explosion.
    Before the second explosion, the separation is the outcome of the previous explosion event, thus the colour coding of panels c and d refers to ranges rather than discrete values.
    Logarithmic scales are used for clarity.}
    \label{fig:v_ainit}
\end{figure*}

If the eccentricity of a given model rises above unity after either explosion (taking evolutionary paths II or III in Figure~\ref{fig:models}), we declare it as an unbound system and measure the peculiar velocities of its components.
Figure~\ref{fig:v_ainit} shows the distributions of the peculiar velocities of the unbound neutron stars and SN progenitors.
Panels a and b show the velocities gained in the first explosion (evolutionary path III), while panels c and d show those gained in the second explosion (path II).
All panels are colour-coded with the separation of the binary prior to the explosions.
In panels c and d ranges of $a_1$ are used, similarly to Figure~\ref{fig:vsys}.
The velocity of both the NS and the SN progenitor star is lower if the separation of the binary prior to the first explosion is higher.
The width of the distributions also decreases with a larger semi-major axis, which means that velocities will generally be lower in initially wider systems.
These trends are slightly smeared out for systems that follow evolutionary path II, because of the non-discrete values of $a_1$, as can be seen on panels c and d of Figure~\ref{fig:v_ainit}.
Multiple maxima can be observed in the distributions of $v_{\mathrm{pri,2}}$ and $v_{\mathrm{sec,2}}$, similarly to the system velocities explored in the previous section, as seen in panels c and d of Figure~\ref{fig:v_ainit}.
Due to the conservation of momentum, the smaller mass component gains a higher peculiar velocity in both explosions.
The less massive component's velocity can be as high as $200~\mathrm{km~s^{-1}}$, while the more massive one leaves the system with an average maximum velocity of about $100~\mathrm{km~s^{-1}}$.
We observe that the velocities of the NSs are about 25-30 per cent smaller if the system takes evolutionary path II rather than III (it dissociates in the second explosion rather than the first).
Evolving through path II means that the velocity of the NS formed in the second explosion is $v_{\mathrm{sec,2}}\leq70~\mathrm{km~s^{-1}}$, and the primary NS gains velocities of $v_{\mathrm{pri,2}}\leq150~\mathrm{km~s^{-1}}$.
The most likely value for both $v_{\mathrm{pri,1}}$ and $v_{\mathrm{pri,2}}$ is about $10~\mathrm{km~s^{-1}}$.
However, the most likely value of $v_{\mathrm{sec,1}}$ is about $12~\mathrm{km~s^{-1}}$, while the most likely value of $v_{\mathrm{sec,2}}$ is only about $7~\mathrm{km~s^{-1}}$.

\begin{figure*}
    \centering
    \includegraphics[width=0.9\textwidth]{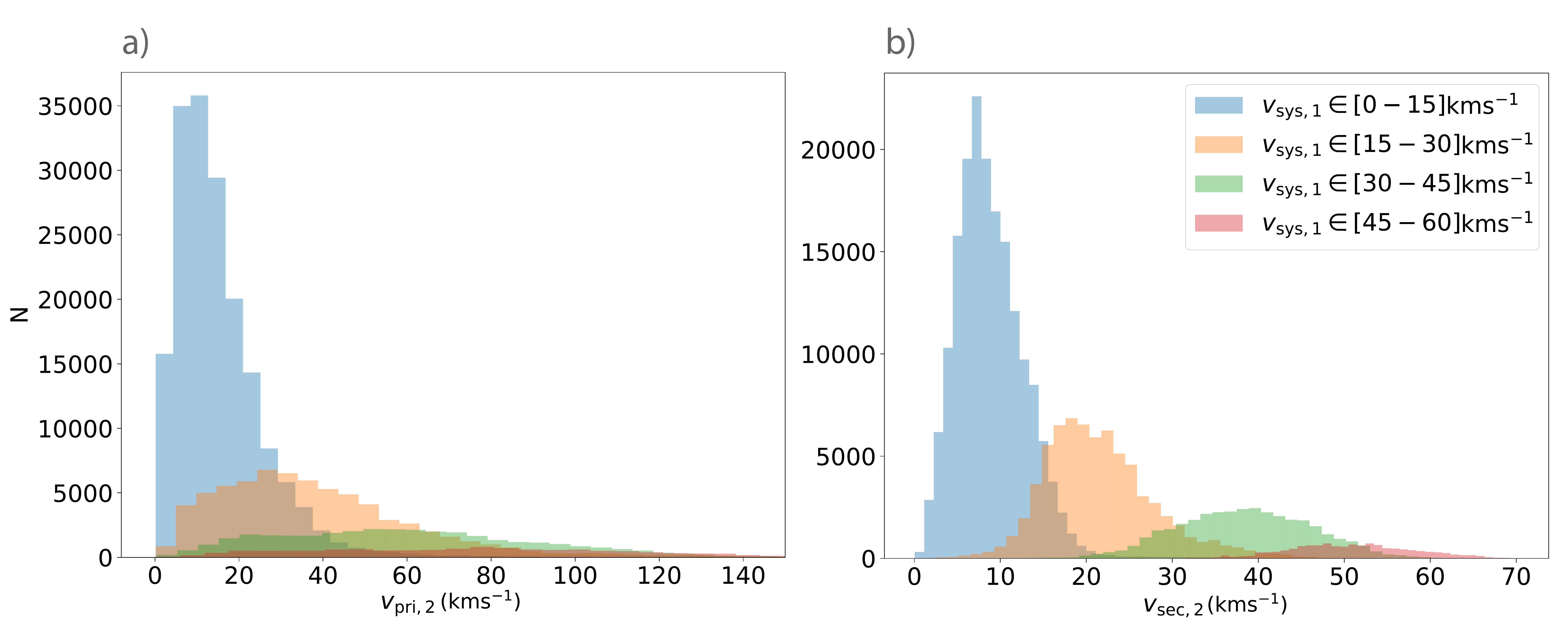}
    \caption{Distribution of NS velocities if the system dissociates during the second explosion.
    Panel a shows the distribution of $v_{\mathrm{pri,2}}$, while panel b shows that of $v_{\mathrm{sec,2}}$. 
    Colours represent different $v_{\mathrm{sys,1}}$ ranges.}
    \label{fig:v_sys_v_sys}
\end{figure*}

Peculiar velocities of the components formed in evolutionary paths II and III do not depend on the true anomaly (neither $\nu_0$ nor $\nu_1$).
A weak trend can be identified in path III with regards to the peculiar velocities as a function of pre-explosion orbital eccentricities.
More circular orbits result in a wider distribution of the peculiar velocities, thus the maximum is shifted towards higher values.
This trend is completely absent in path III.


Figure~\ref{fig:v_sys_v_sys} shows the peculiar velocities of the neutron stars following path II, colour-coded with the system velocities gained in the first explosion.
The velocities of both NS-s are proportional to the system velocity $v_{\mathrm{sys,1}}$.
As seen in panel a, the width of the distribution of the primary NS's velocity, $v_{\mathrm{pri,2}}$, increases with $v_{\mathrm{sys,1}}$, and the most probable value shifts toward higher velocities.
The width of the distribution of $v_{\mathrm{sec,2}}$, however, changes only slightly with $v_{\mathrm{sys,1}}$, while the most probable value of velocity also increases (see panel b).
With regard to the minimum value of NS velocities, we find that the minimum of $v_{\mathrm{pri,2}}$ is 0.13~$\mathrm{km~s^{-1}}$.
However, the minimum of $v_{\mathrm{pri,2}}$ is proportional to $v_{\mathrm{sys,1}}$.

\section{Discussion}
\label{sec:discussion}

\subsection{Weighted probabilities}

Not all binary configurations are equally likely at the moment of explosion in the parameter space studied.
To account for the occurrence rates of different systems, we calculate weighting factors of the true anomaly and the mass ratio (see Appendix~\ref{appendix} for a detailed description of the calculations).
We then compute new histograms with the final weighting factor, $w$,  thus obtaining weighted distributions of the data.
Note that $w$ is obtained as the product of the true anomaly and mass ratio weighting factors, which are assumed to be independent.

\begin{figure}
    \centering
    \includegraphics[width=\columnwidth]{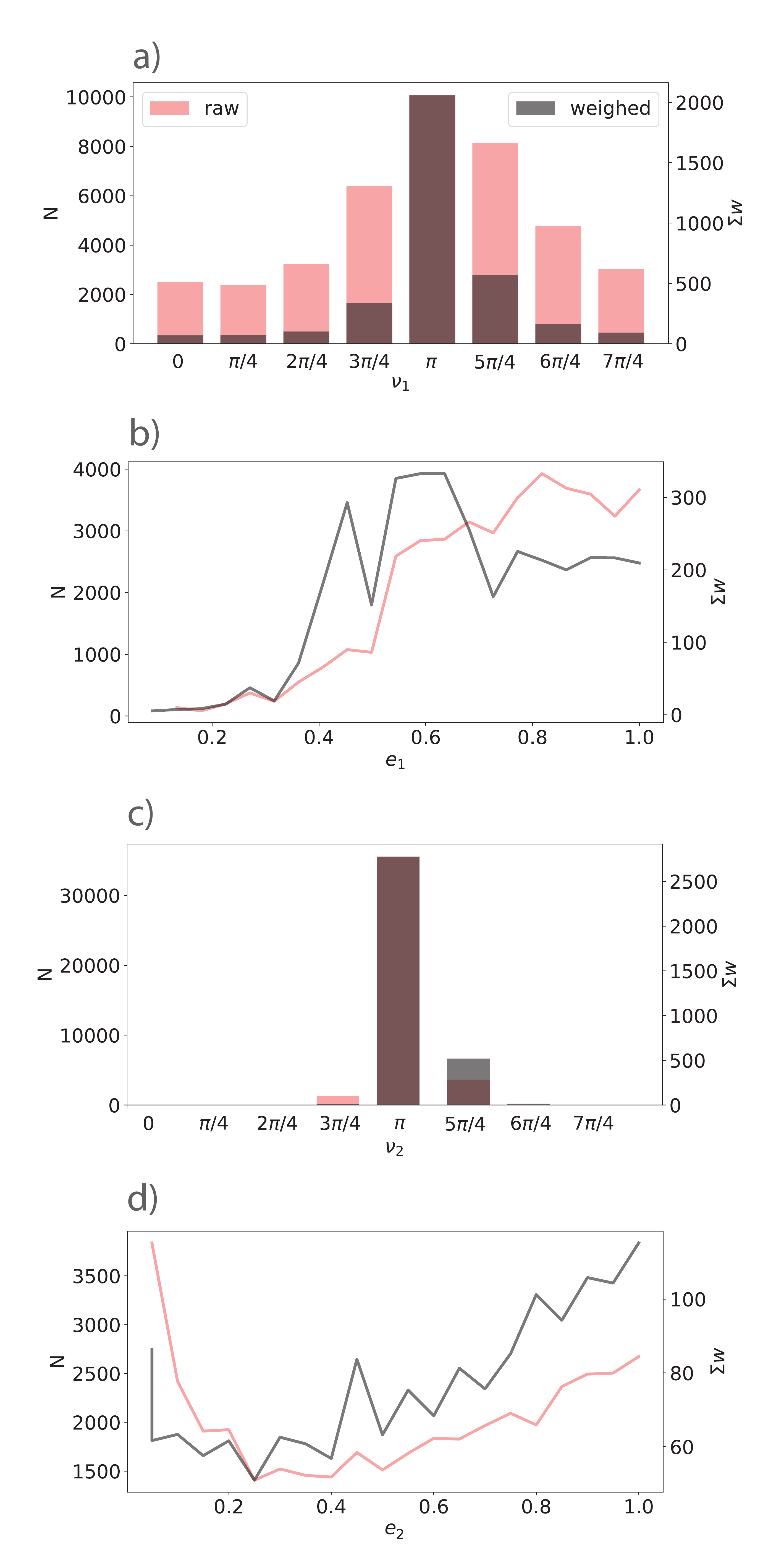}
    \caption{Distribution of $\nu_1$, $e_1$, $\nu_2$ and $e_2$ of DNS systems where weighing alters the trends shown in Section \ref{sec:results}.
    The raw data without weighing are shown in light red, while grey shows the distribution of the data after weighing.}
    \label{fig:w}
\end{figure}

Panels a-d of Figure~\ref{fig:w} show the distributions of $\nu_1$, $e_1$, $\nu_2$ and $e_2$ of the DNS systems.
Weighing does not alter the trends shown in Section~\ref{sec:results} for the other parameters examined.
Light red shows the raw data, while the distribution of the weighted data is shown in grey.
As expected, the true anomaly values closer to the apocentre are given significantly more weight, while the asymmetry shown in Figure~\ref{fig:nu} is maintained (see panels a and c of Figure~\ref{fig:w}).
On panel b one can see that the probability of highly eccentric ($e_1>0.7$) DNS progenitor systems decreases in favour of mildly eccentric NS-SN progenitor binaries ($0.4<e_1<0.7$).
A weaker trend can be seen in panel d with respect to the eccentricity of DNS systems, where models with low eccentricities ($e_2<0.1$) are less likely to occur if the data are weighed.
Otherwise, the distribution of DNS eccentricities follows a quasi-linear function, with higher eccentricity models being more favoured after weighting.
In terms of the occurrence rates of different evolutionary paths, weighing does not significantly alter the results.
The probabilities after weighting are as follows: 5.6 per cent for evolutionary path I, 36.4 per cent for path II, and 58 per cent for path III.

\subsection{Orbital parameters of DNS systems}

According to our simulations, the minimum separation of a DNS system is 2.95 au. 
We can thus confirm that the observations of DNS systems in tight orbits require at least one stage of mass transfer during the lifetime of the system \citep{podsiadlowski-etal-1992}, a common envelope phase \citep{paczynski-1976, ostriker-1976}, or some other orbit-shrinking mechanism.

In the parameter space we studied, the stability of the models in the first - and in the second - explosion is insensitive to the separation before the explosion.
In their study, \citet{tauris-etal-2017} found that binaries wide enough to avoid mass transfer are likely to break up in the first SN.
Note, however, that in their study mass loss is modelled as instantaneous and the SN explosion is assumed to be asymmetric.
    
With regard to the shrinking of the orbit in the second explosion, we find contradictory results:
according to our previous findings, the separation of a binary always increases during an SN~II explosion \citep{regaly-etal-2022}.
Since the eccentricity is not pre–defined preceding the second explosion (rather it is an outcome of the first simulation), it can be higher than the pre–defined values examined in the first explosion ($e_1>0.8$ is possible, while $e_0$ is always $\leq0.8$).
However, modelling the first explosion with an additional 1920 systems whose eccentricity is $e_0>0.8$ validate the widening of the orbit.
These extremely eccentric shrinking binaries require further studies.
Note that, however, \citet{tauris-etal-2017} also posed that orbit shrinking is possible.

Systems with arbitrary $\nu_0$ values can survive the first explosion. However, $\nu_1\in[3\pi/4; 6\pi/4]$ is required to avoid dissociation after the second explosion.
For both explosions, the maximum number of bound models is at the apocentre (see Figure~\ref{fig:nu}).
As it has been shown by \citet{regaly-etal-2022}, orbital stability of a planetary-mass companion strictly requires that it is at apocentre, meaning an extremely narrow distribution of true anomaly.
Thus, the much narrower distribution of $\nu_1$ compared to $\nu_0$ can be explained by the fact that the mass of the primary neutron star is small ($\leq 3M_\odot$) compared to that of the exploding secondary.
Note that the distributions are asymmetric towards higher values, which is a consequence of the fact that the true anomaly value corresponds to the onset of the explosion.
Thus, the perturbation of the secondary's orbit starts slightly later, when the expanding envelope reaches the secondary, see details in \citet{regaly-etal-2022}.

According to our simulations, the eccentricity of 25 per cent of the DNS systems is above 0.8, which is similar to the results of \citep{charuasia-bailes-2005}, who state that about 50 per cent of the DNS systems have eccentricities greater than 0.8. 
This also implies that high eccentricity orbits could be tracers of a recent SN explosion \citep{srinivasan-vdheuvel-1982}.
With regard to the observation of colliding DNS systems, a highly eccentric orbit is preferable, as more eccentric orbits lead to quicker orbital decay.

Orbital circularisation during the first explosion is observed in 17 per cent of models.
However, in the second explosion, circularisation is more prominent, 77 per cent of models show a decline in orbital eccentricity.
This results in a wider distribution of eccentricities for DNS systems compared to the distribution of $e_1$, see Figure~\ref{fig:e}.
With regard to the orbital circularisation of DNS systems, we find that $e_1\geq0.36$ and $\nu_1=[\pi;5\pi/4]$ are required.
This is in agreement with our previous finding \citep{regaly-etal-2022} that circularisation of binary star systems is only possible if $e\geq0.4$.

\begin{figure}
    \centering
    \includegraphics[width=\columnwidth]{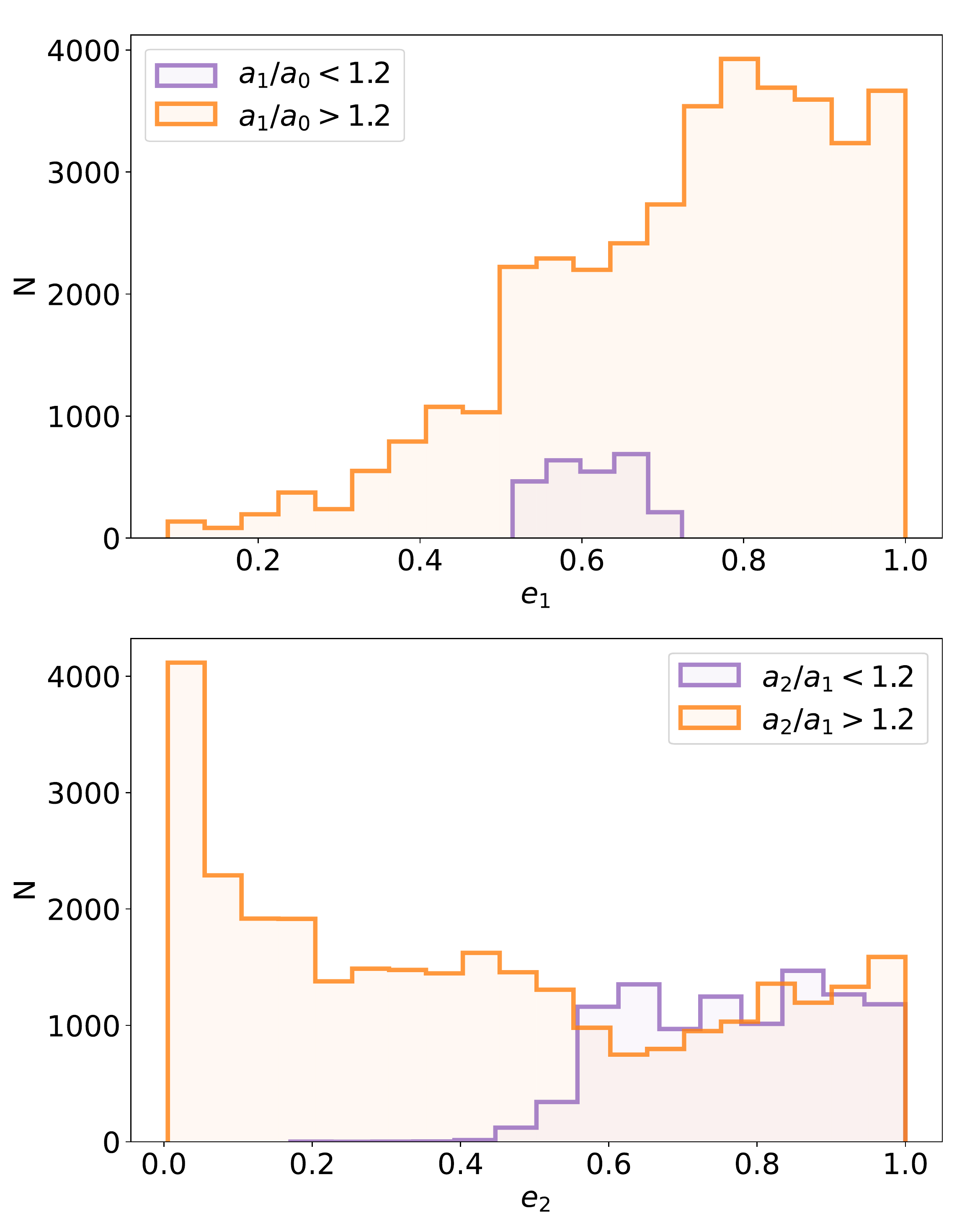}
    \caption{ Distributions of system eccentricities after the first (top panel) and second (bottom panel) explosions.
    The colours represent different regimes of orbit widening caused by the SN explosion.}
    \label{fig:e_agrowth}
\end{figure}

Figure \ref{fig:e_agrowth} shows the eccentricity distributions of the systems after the first and second explosions. 
The two colours represent two different regimes of orbital widening. 
In the upper panel we can see that when $a_1/a_0<1.2$, the eccentricity of the NS-SN progenitor system is always between 0.5 and 0.7. 
However, if $a_1/a_0>1.2$, $e_1$ can be arbitrary. 
The second explosion also gives similar distributions: 
if $a_2/a_1<1.2$, $e_2>0.4$, and if $a_2/a_1>1.2$, $e_2$ can have any value. 
This contradicts the results of \citet{kalogera-1996}, who states that the eccentricity of DNS systems is low when $a_2/a_1\simeq1$. 
Note, however, that \citet{kalogera-1996} assumed an instantaneous mass loss to model the effect of NS kicks, while in our model, mass loss and orbit perturbation occur over a longer period.

\subsection{DNS system velocities and high peculiar speed pulsars}

With regard to the dependence of system velocities on ejecta masses, $v_{\mathrm{sys,1}}$ is independent of the mass lost in the first explosion (see the overlapping distributions on panel c of Figure ~\ref{fig:vsys}).
However, the distributions of $v_{\mathrm{sys,2}}$ have non-overlapping regions (panel d of Figure ~\ref{fig:vsys}), thus the velocity of DNS systems is weakly dependent on the mass ejected in the second explosion.
This can be explained by comparing the mass lost to the total mass of the system.
In the second explosion, $Q_2=M_{\mathrm{ej,2}}/(M_{\mathrm{n,1}}+M_{\mathrm{ej,2}}+M_{\mathrm{n,2}}) \in [0.52-0.88]$, while this ratio in the first explosion is smaller,  $Q_1=M_{\mathrm{ej,1}}/(M_{\mathrm{ej,1}}+M_{\mathrm{n,1}}+M_{\mathrm{ej,2}}+M_{\mathrm{n,2}}) \in [0.34-0.7]$.
Thus, as the systems obey the conservation of momentum, the system velocity is more sensitive to the ratio of mass loss in the second explosion.
 
\begin{figure}
    \centering
    \includegraphics[width=\columnwidth]{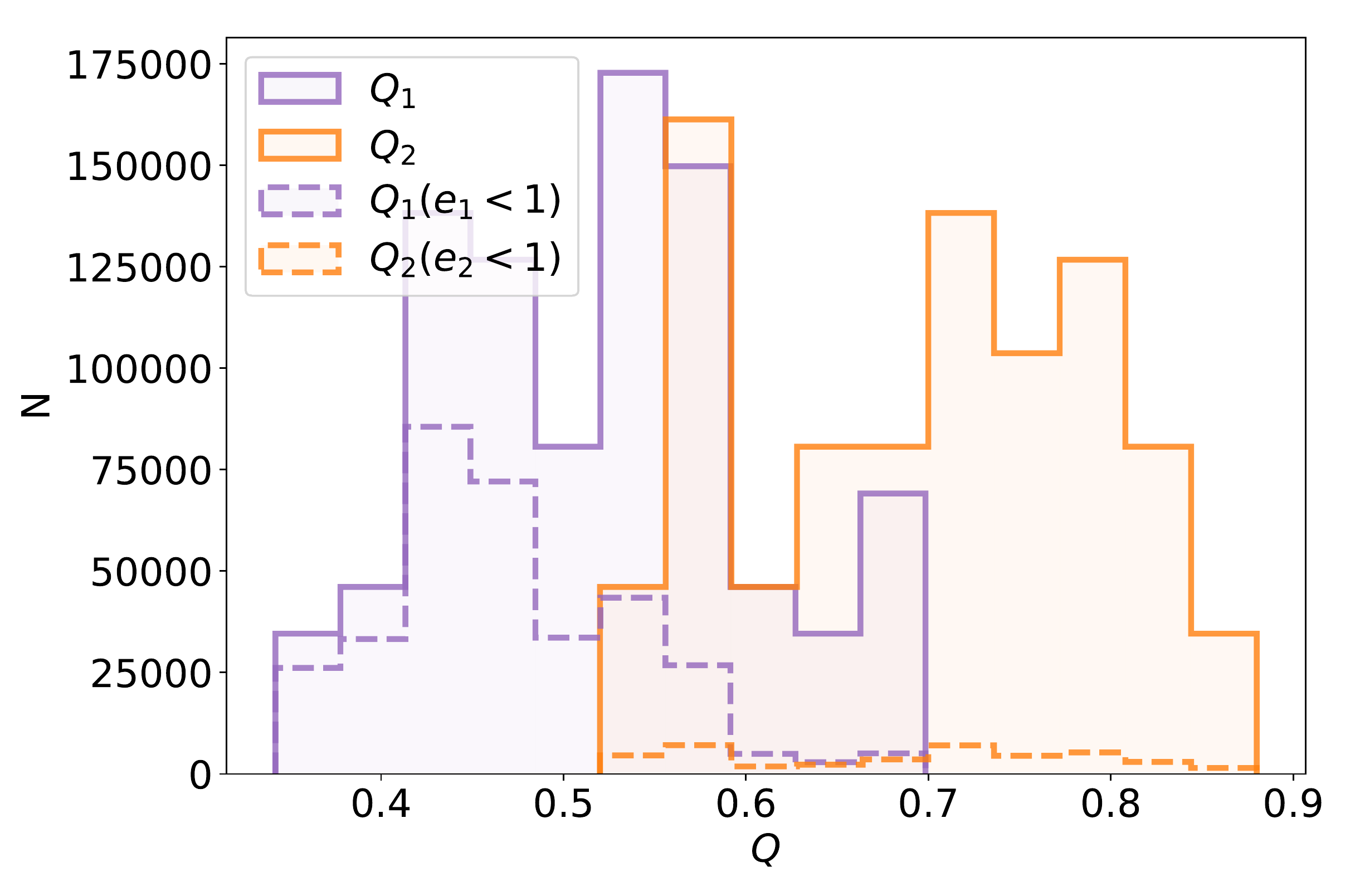}
    \caption{Distributions of ratios of ejecta mass to total system mass before explosions.
    The histograms with solid lines show all systems with a given $Q$ ratio, while the histograms with dashed lines show only the stable systems.}
    \label{fig:e_Q}
\end{figure}

Figure~\ref{fig:e_Q} shows the distributions of the ratio of mass lost in each explosion to the total mass of the system, i.e. $Q_1$ and $Q_2$. 
Also shown, with dashed lines, are the histograms of the models that survive each explosion on a stable orbit.
It can clearly be seen that during the first explosion a smaller $Q_1$ is slightly preferred for stability, but during the second explosion $Q_2$ can be arbitrary.
Note that more than half of the mass of the system is always lost in the second explosion.
It has been shown by many authors (see e.g. \citealp{dewey-cordes-1987, hills-1983, veras-etal-2011}) that the ratio of mass loss during a SN explosion determines the orbital stability of the system.
\citet{hills-1983} states that binaries can survive over 50 per cent mass loss if their orbit is eccentric and the stars are at the apocentre at the moment of explosion.
As noted in section~\ref{sec:results}, the formation of DNS systems by subsequent SN~II explosions also requires the stars to be at the apocentre of an eccentric orbit.
If $Q$ is defined as $M_{\mathrm{ej}}/(M_{\mathrm{ej}}+M_{\mathrm{n}})$, i.e. the mass loss is only compared to the mass of the exploding star, such trends cannot be seen. 
This alternative ratio is always $\geq0.7$, and an arbitrary value can lead to the formation of a stable system.

In our previous study \citep{regaly-etal-2022}, we hypothesised a possible doubling of the system velocity of bound systems during the second SN~II explosion. 
This, however, did not happen, as the system velocities in both explosions are in the same range, shown in Figure~\ref{fig:vsys}.
This is explained by the fact that doubling the system velocity requires a very special configuration, in which both system velocity vectors are parallel and point in the same direction.
The average system velocity of the DNS systems examined in this study is $17~\mathrm{kms^{-1}}$. 
According to \citet{tauris-etal-2017}, the mean DNS system velocity is $30~\mathrm{kms^{-1}}$. 
For an asymmetric explosion, the mean could be as high as $100~\mathrm{kms^{-1}}$ \citep{kornilov-lipunov-1983, dewey-cordes-1987}.
The discrepancy between our predicted and the observed system velocities can be explained by the fact that our model neglects the possibility of an asymmetric explosion.

The peculiar velocity of the NS-s following evolutionary path II, $v_{\mathrm{pri,2}}$ and $v_{\mathrm{sec,2}}$, are strongly dependent on the system velocity gained in the first explosion, $v_{\mathrm{sys,1}}$, as shown by the non-overlapping distributions colour-coded with different $v_{\mathrm{sys,1}}$ ranges in Figure~\ref{fig:v_sys_v_sys}.
Note that $v_{\mathrm{sys,1}}$ in turn is determined by the initial separation of the binary, $a_0$ as it is shown in Figure~\ref{fig:v_ainit}.
The velocity of the secondary NS ($v_{\mathrm{sec,2}}$) is more affected by the system velocity compared to the primary NS ($v_{\mathrm{pri,2}}$), as can be seen in Figure~\ref{fig:v_sys_v_sys}.
On the one hand, this is because the velocity of the secondary is not affected by its expanding shell during the second explosion, and thus inherits the previously gained system velocity.
On the other hand, the velocity distribution of the primary NS is spread out by the velocity of the expanding shell of the secondary NS.

We have seen in Section~\ref{sec:results} that the velocities of the NSs are about 25-30 per cent lower when the system takes evolutionary path II rather than III.
This is in agreement with \citet{tauris-etal-2017}, who state that the second kick received by the NS is smaller than the first.

According to our simulations, the maximum PSR peculiar velocity is $275~\mathrm{kms^{-1}}$. 
This contradicts the finding of \citet{lyne-lorimer-1994} that the average peculiar velocity of pulsars in the Galaxy is $450~\mathrm{kms^{-1}}$.
Since the Galactic escape velocity is $550~\mathrm{kms^{-1}}$ \citep{smith-etal-2007}, a fraction of these high peculiar velocity pulsars may even leave the Galaxy after the SN explosion.
Assuming a symmetric explosion, peculiar velocities $>275~\mathrm{kms^{-1}}$ cannot be achieved, even if the expansion velocity is above 10~000$\mathrm{kms^{-1}}$. 
However, the contradiction can be resolved by an asymmetric SN explosion.

An asymmetry in the mass distribution of a star can be a consequence of Roche--lobe filling, which is likely to lead to type Ib/c SN explosions.
According to \citet{tauris-vdheuvel-2006}, Roche--lobe filling can occur for binary periods of up to 10 years. 
This corresponds to separations of $a_0 \in [11-18]~\mathrm{au}$ in our studied parameter space, depending on the masses of the stars. 
Thus, for models with an initial separation of $a_0\leq10~\mathrm{au}$, the envelope asymmetry caused by the Roche--lobe overflow should not be neglected.

\section{Conclusions}
\label{sec:conclusions}

In this study, we presented a consecutive SN~II formation channel of double neutron star (DNS) systems.
In our model these systems form via two subsequent SN~II explosions.
We simulated both explosions with a homologous expansion model (see a detailed description in \citealp{regaly-etal-2022}) and monitored the change of the orbital parameters and velocities using a numerical 8th-order explicit Runge--Kutta integrator.
The simulated systems can take three different evolutionary paths (denoted with I, II and III), as summarised in Figure~\ref{fig:models}.
The majority of the investigated systems (63 per cent of the total 1~658~880 simulations) dissociated, i.e. they took the evolutionary path II or III.
37 per cent of systems remained bound after the first explosion, and only 4.5 per cent of all systems were able to keep stability after the secondary star exploded, and have formed DNS systems (path I).
Having analysed the outcomes of the simulations, our major conclusions are:

1) The orbit of the examined systems always expands during the first explosion (as expected) and is generally further widens after the second explosion.
Regarding the first explosion, the post-SN separation of the binary is most likely 3-4 times larger than the pre-SN separation.
On the contrary, this ratio in the second explosion is almost always smaller than four.
Systems that remain bound after the double SN~II explosion have a minimum separation of 2.95~au.

2) In some cases (204 models, which comprise only 0.01 per cent of all investigated initial conditions), the second explosion can cause the orbit to shrink. 
This, however, requires a very special fine-tuning of the initial conditions: a slowly expanding shell ($v_{\mathrm{max}}=1000\mathrm{km~s^{-1}}$), and a highly eccentric ($e_1>0.88$) secondary residing at the apocentre ($\nu_1=\pi$) before the second explosion. 

3) During the first explosion, the eccentricity of systems increases if $e_0<0.8$ .
However, 75 per cent of models assuming $e_0=0.8$ circularise.
Almost all systems where circularisation happens reside exactly at the apocentre at the moment of the first explosion.

4) Orbits of models where $e_1\geq$0.4 circularise in the second explosion, which comprises the majority of surviving models. 
Thus the distribution of the eccentricities becomes more homogeneous compared to the first explosion.
However, eccentricities increase during the second SN explosion if $e_1$<0.4.

5) The formation of a DNS system (evolutionary path I) requires the secondary to reside close to the apocentre at the moment of the second explosion ($\nu_1 \approx \pi$).
If the secondary is apart from the apocentre, the orbital eccentricity of the DNS system increases significantly.
The systems take evolutionary path II and dissociate during the second explosion if $\nu_1 \notin [3/4\pi; 6/4\pi]$.

6) DNS system velocities are found to be less than 60~$\mathrm{km~s^{-1}}$, and are only weakly dependent on ejecta mass.

7) The tighter the pre-explosion orbit of the stars, the higher the peculiar velocity. However, ejecta masses do not affect the peculiar velocities of the dissociating systems' components. The component velocities of systems following evolutionary path III are in the range of 1.4-234~$\mathrm{km~s^{-1}}$, while those following path~II are in the range of 0.02-150~$\mathrm{km~s^{-1}}$.

Finally, let us emphasise our novel results.
The presented model can explain the formation of wide-orbit (>2.95~au) DNS systems in the investigated parameter regimes.
The initial separation of the systems could not be set lower than $a_0=R_0$, because this value represents a contact binary system for the progenitor stars.
Detection of gravitational waves requires a DNS system tighter than 2.95~au, because gravitational wave emission with our current instruments can only be detected when the objects are only a few radii apart. 
Thus, the scenario presented here can only produce systems that emit observable gravitational waves if orbital shrinking occurs after the formation of the DNS system.
This handicap is similar to the final parsec problem of supermassive black hole binary mergers (see \citealp{milosavljevic-merritt-2003} for a summary, and \citealp{berczik-etal-2006} for a possible solution).
According to our homologous expansion model, the envelope mass within 10~au can be as high as $1M_\odot$ after five million days.
Conventional type II migration can occur in such a gaseous medium \citep{LinPapaloizou1986}.
However, we have to take into account the non-negligible system velocity, which is different from the velocity inherited by the two expanding shells.
Exploring this hypothesis requires further combined N-body and hydrodynamical simulations.

The system velocities of the DNS systems observed so far fall in the range of 28-240~$\mathrm{km~s^{-1}}$, the vast majority of these being smaller than 60~$\mathrm{km~s^{-1}}$ \citep{tauris-etal-2017}.
Thus our spherically symmetric expansion model can explain most of the observed DNS velocities.
However, the explanation of PSR B1534+12 \citep{fonseca-etal-2014} and PSR B1913+16 \citep{hulse-taylor-1975}, whose system velocities are 138, and 240~$\mathrm{km~s^{-1}}$, respectively, requires a more elaborate expansion model in which either of the explosions can be asymmetric, or a completely different DNS formation scenario happens.

\section*{Acknowledgements} 

JV is supported by the project OTKA-K142534 of the National Research, Development, and Innovation Office (NKFIH), Hungary.
VF acknowledges financial support from the ESA PRODEX contract nr. 4000132054.
We thank Cs. Kalup for a fruitful discussion on the Gaia biases.
We thank the anonymous referee, whose helpful suggestions greatly improved the quality of our paper. 


\section*{Data availability}
The data underlying this article obtained with the Homologous Expansion Python code (HEPy) can be shared at a reasonable request to the corresponding author.




\appendix
\section{Calculation of weighing factors}
\label{appendix}

Since not all initial binary configurations are equally likely to occur, weighted probabilities are calculated based on the true anomaly and the initial mass ratio of the binary.  
We have seen in Section \ref{sec:results} that the remnant mass has a negligible quantitative effect on the results of the simulations, which is in agreement with \citet{ozel-etal-2012}.
Therefore, the models are not weighted by the NS mass.

Different orbital positions are not occupied for the same amount of time, i.e. the components on an eccentric orbit are more likely to be close to the apocentre when the SN explosion occurs. 
To obtain the true anomaly weighting factor $w_\nu$, the time interval that the system spends at a $\nu \pm \pi /8$ section of the orbit is calculated and compared to the entire binary period.
The period of a binary is
$P = 2\pi a^{3/2} / \sqrt{\mu}$.
The eccentric anomaly can be calculated as
\begin{equation}
    E = 2 \arctan{
    \left( 
    \sqrt{ \frac{ 1-e }{ 1+e } }
    \tan{ 
    \left(
    \frac{\nu \pm \pi/8}{2} 
    \right)
    } 
    \right) 
    },
\end{equation}
The mean anomaly is the result of the solution of the Kepler equation of the system:
$M = E - e\sin(E)$.
The time elapsed since passing the pericentre, $\tau$, can then be calculated as
$\tau = M P / (2\pi)$,
and thus the time elapsed between true anomaly values $\nu\pm \pi/8$ is obtained as
$\Delta t = \tau(\nu+\pi/8)-\tau(\nu-\pi/8)$.
The true anomaly weighting factor is $w_{\nu}=\Delta t/P$.

We use the statistics of stellar systems with different primary masses, mass ratios and periods given in \citet{moe-distefano-2017} to derive a mass ratio weighting factor based on their Table~13, which presents the shape of the probability density function (PDF) as a function of the above properties.
Defining $q$ as the initial mass ratio $(M_{\mathrm{ej,pri}}+M_{\mathrm{n,pri}})/(M_{\mathrm{ej,sec}}+M_{\mathrm{n,sec}})$, the PDF can be divided into three regimes (see Figure 2. in \citealp{moe-distefano-2017}):
$i)~0.1<q<0.3$, $ii)~q>0.3$, and $iii)~q>0.95$.
The first two regimes are defined by power laws, $q^{\gamma_{\mathrm{smallq}}}$ and $q^{\gamma_{\mathrm{largeq}}}$, respectively.
The last regime defines an excess probability fraction of binaries whose components are nearly equal in mass. 
In our case this excess fraction is always $<0.03$, so we can safely ignore it.
To construct a PDF it is necessary that 
\begin{equation}
    \alpha \int_{0.1}^{0.3} q^{\gamma_{\mathrm{smallq}}}+\beta \int_{0.3}^{1} q^{\gamma_{\mathrm{largeq}}}=1,
\end{equation}
where $\alpha$ and $\beta$ are constants.
Since the PDF is a continuous function, i.e. $\alpha~ 0.3^{\gamma_{\mathrm{smallq}}}=\beta ~0.3^{\gamma_{\mathrm{largeq}}}$, $\beta$ can be derived after some trivial algebra as
\begin{equation}
    \beta = \frac{1}{m~0.3^{\gamma_{\mathrm{largeq}}-\gamma_{\mathrm{smallq}}}+n},
\end{equation}
where 
\begin{equation}
m=\int_{0.1}^{0.3}q^{\gamma_{\mathrm{smallq}}}dq
\end{equation}
and 
\begin{equation}
n=\int_{0.3}^{1}q^{\gamma_{\mathrm{largeq}}}dq.
\end{equation}
In our case, all models are $q>0.3$. Therefore, the mass ratio weighting factor, $w_\mathrm{q}$, can be expressed as
\begin{equation}
    w_{\mathrm{q}} = \beta q^{\gamma_{\mathrm{largeq}}} = \frac{q^{\gamma_{\mathrm{largeq}}}}{m0.3^{\gamma_{\mathrm{largeq}}-\gamma_{\mathrm{smallq}}}+n}.
\end{equation}
The exponent of the power law, $\gamma_{\mathrm{largeq}}$, depends on the binary period, see Table~13 of \citet{moe-distefano-2017}.
We consider the weighting factors to be independent, so $w=w_{\nu_1} w_{\mathrm{q}}$ for the first explosion and $w=w_{\nu_1} w_{\nu_2} w_{\mathrm{q}}$ for the second explosion.

\bsp	
\label{lastpage}
\end{document}